\documentclass[12pt]{article}
\usepackage[english]{babel}
\usepackage[utf8]{inputenc}
\usepackage[T1]{fontenc}

\def\bAi{(\bA^{-1})}

\usepackage{amsmath}

\usepackage{amsfonts}

\usepackage{url}
\usepackage[dvips]{graphicx}
\usepackage{calc}
\usepackage{subfigure}
\usepackage{multirow}

\def\mL{\mathcal{L}}
\def\mH{\mathcal{H}}

\def\bA{\mathbf{A}}

\def\bD{\mathbf{D}}

\begin{document}
	\begin{titlepage}
		\begin{center}
			{\Large{ \bf Covariant Canonical Formalism For Born-Infeld Inspired Gravity in Palatini Formulation}}
			
			\vspace{1em}  
			
			\vspace{1em} J. Kluso\v{n} 			
			\footnote{Email addresses:
				klu@physics.muni.cz  }\\
			\vspace{1em}
			\textit{Department of Theoretical Physics and
				Astrophysics, Faculty of Science,\\
				Masaryk University, Kotl\'a\v{r}sk\'a 2, 611 37, Brno, Czech Republic}
			
			\vskip 0.8cm
			
			%
			%
			%
			%
			%
			%
			
			\vskip 0.8cm
			
		\end{center}

		\begin{abstract}
		We analyze Born-Infeld inspired gravity in Palatini formulation in the framework of covariant canonical formalism. We determine covariant Hamiltonian and corresponding equations of motion. 	
		\end{abstract}
		
		\bigskip
		
	\end{titlepage}
	
	\newpage

\section{Introduction and Summary}\label{first}
Usually the gravitational equations of motion are derived from Einstein-Hilbert action where the metric is treated as dynamical 
variable. We obtain these equations by variation of the 
action with respect to metric even if strictly speaking this variation is not well defined due to the presence of the second derivative of the metric in Ricci scalar. More preciselly, the variation of Einstein-Hilbert action is correctly defined when we include an appropriate boundary term to the action which depends on the choice of the boundary surface. Even if these boundary terms do not affect equations of motion they are very important for thermodynamics aspects of gravity 
\cite{Chakraborty:2019doh,Padmanabhan:2013nxa}. Then there is natural to ask a question whether it is possible to formulate gravity action with well defined variation principle. One such an intriguing proposal was formulated almost 100 years ago by 
Eddington \cite{Eddington} which was recently reconsidered in 
\cite{Banados:2010ix}. In this approach the Lagrangian for gravitational field is given by expression $\sqrt{\det R_{ab}}$ where $R_{ab}$ is Ricci tensor that can be defined using the connection $\Gamma^c_{ab}$ only which means that $\Gamma^c_{ab}$ is dynamical variable whose equations of motion are derived by variation of the action with respect to them. Interestingly the
resulting equations are Einstein equations with non-zero cosmological constant. 

However the major issue with Eddington gravity is that it does not contain coupling to matter. Then in order to include matter it was proposed that this theory should be extended with metric so that the resulting action has Born-Infeld like structure
\cite{Born:1934gh}. First attempt to implement Born-Infeld like structure for gravity was performed in 
\cite{Deser:1998rj} where the fundamental degrees of freedom were components of metric. However since in this case action for Born-Infeld gravity contains higher order derivatives of  metric it was quickly recognized that this theory suffers from Ostrogradski instability 
associated to higher order equations of motion. In order to avoid problems with ghost we should consider connection as an independent variable \cite{Vollick:2003qp}. Due to the fact that now Ricci tensor contains first order derivatives of connections the ghost problem 
is imediatelly solved. These ideas were further elaborated 
for example in \cite{Banados:2010ix,Vollick:2005gc}
and these theories are now known as Born-Infeld like inspired theories of gravity , for review and extensive list of references, see 
\cite{BeltranJimenez:2017doy}. Due to the fact that these Born-Infled like inspired theories of gravity on the one side are free from ghost and on the other side they significantly differ from standard general relativity when scalar curvature is of order of the mass scale of new physics, these theories have been studying in many papers up to present time. 

One of the most interesting question related to BI gravity  is its canonical analysis. This is difficult task  since in $3+1$ formalism  manifest covariance of the theory is lost
\cite{Gourgoulhon:2007ue}. Moreover, in case of BI gravity where the metric is dynamical degrees of freedom  
we have to deal with the fact that in $3+1$ formalism 
there are terms in the action with second derivative of metric. Then in order to have an action with first order derivatives only we should introduce new degrees of freedom. In case of Palatini formalism corresponding formulation is also very complicated and has not been performed yet again due to the lost of manifest covariance of theory. For that reason it would be desarable to have treatment that on hand leads to canonical formulation and on the other hand preserves manifest covariance of theory. This approach was formulated long time ago and it is known as Weyl-De Donder formalism \cite{DeDonder,Weyl}.
 The key point of this formulation is that now canonical Hamiltonian
density depends on conjugate momenta $p^\alpha_M$ which are variables conjugate to $\partial_\alpha \phi^M$. In other words we tread all partial derivatives on the equal footing which clearly preserves diffeomorphism invariance. This approach is known as
multisymplectic field theory, see for example
\cite{Struckmeier:2008zz,Kanatchikov:1997wp,Forger:2002ak}, for review, see \cite{Kastrup:1982qq}.

In \cite{Kluson:2024jga}
we  used this formalism for BI gravity where metric components are dynamical degrees of freedom and we showed that in this case the theory possesses additional degrees of freedom which confirms the result found in 
\cite{Deser:1998rj}. In this paper we would like to apply covariant canonical formalism to the case of Palatini formulation of Born-Infeld gravity. We find that in this case the transformation from Lagrangian to Hamiltonian formulation is more straightforward. Then we determine canonical equations of motion and analyze their properties firstly in case of absence of matter and then when the matter content is non-zero. We study their solutions and   we show that they are equivalent to solutions of Lagrangian equations of motion. We mean that this is nice consistency check of covariant canonical formalism. We also show that equations of motion of BI gravity significantly differ from 
equations of motion of general relativity in the regime where the curvature is of order of the scale that defines BI gravity. 

Despite of the fact that covariant canonical formalism provides elegant description of the system there is still problem with practical application of this formalism. In case of general relativity covariant canonical formalism has nice relation to thermodynamics quantities due to the presence of boundary term in the action 
\cite{Parattu:2013gwa,Padmanabhan}. In case of Born-Infeld gravity 
the situation is different since due to the fact that the variational principle is well defined corresponding boundary term is missing. In fact, as far as we know the thermodymics character od Born-Infeld gravity is not well understood and maybe covariant canonical formulation of Born-Infeld gravity could be useful for 
its development.  

The organization of this paper is as follows. In the next section (\ref{second}) we review basic properties of Born-Infeld inspired gravity in Palatini formalism and determine  equations of motion. In section (\ref{third}) we find corresponding covariant Hamiltonian. Then we determine canonical equations of motion and discuss their properties firstly in the case of absence of matter and then when the matter is present. We show in both cases that 
the resulting equations of motion are equivalent to Lagrangian ones.

\section{Review of BI gravity in  Palatini formalism}\label{second}
Let us consider Born-Infeld (BI) inspired  gravity action where the metric and connection are independent
\begin{equation}\label{actBI}
	S=M^2_pM^2_{BI}\int d^4 x [\sqrt{-\det \bA}-\lambda\sqrt{-g}]+S_M	\ , 
\end{equation}
where 
\begin{equation}
	\bA_{ab}=g_{ab}+\frac{1}{2M_{BI}^2}(R_{ab}+R_{ba}) \ , \quad 
\end{equation}
where
\begin{equation}
	R_{ab}=\partial_m \Gamma^m_{ab}-\partial_a \Gamma^m_{bm}
	+\Gamma^m_{mn}\Gamma^n_{ab}-\Gamma^m_{an}\Gamma^n_{mb} \ , 
	\quad a,b,c\dots,=0,1,2,3 \ , 
\end{equation}
and where $S_M$ is matter part of action. Note that the most general form of BI gravity contains connection with torsion
\footnote{For detailed discussion the most general cases see for example review \cite{BeltranJimenez:2017doy}.}
		 however we restrict ourselves to the case of torsion-free connection 
which means that connection is symmetric
\begin{equation}
	\Gamma^c_{ab}=\Gamma^c_{ba} \ . 
\end{equation}
Further,  $M_{BI}$ is the mass scale which determines when higher order corrections are important. We should also stress that  the last term proportional to $\lambda$ is necessary in order to remove 
potential cosmological constant in order to make Minkowski space-time as solution of BI gravity. In fact, cosmological constant can be determined by expansion of the square root form of the action in higher orders of $R$. Then term proportional to $\sqrt{-g}$ corresponds to the $2\Lambda$ where the cosmological constant $\Lambda$ is equal to 
\begin{equation}
	\Lambda=(\lambda-1)M^2_{BI} \ . 
\end{equation}
Finally  $g_{ab}$ is four dimensional metric  where however  $\Gamma_{ab}^c$ and $g_{ab}$ are independent in Palatini formalism.  

It is instructive to determine equations of motion that follow from (\ref{actBI}). Since $g_{ab}$ and $\Gamma^c_{ab}$ are independent, the  variation of the action has the form 
\begin{eqnarray}
&&	\delta S=\frac{1}{2}M_p^2M_{BI}^2
	\int d^4x [\sqrt{-\det\bA}\bAi^{ab}(\delta g_{ab}+\frac{1}{M_{BI}^2}\delta R_{(ab)})-\nonumber \\
&&	-\lambda\sqrt{-\det g}
	g^{ab}\delta g_{ab}]+\frac{\delta S_M}{\delta g_{ab}}\delta g_{ab} \ , \nonumber \\
\end{eqnarray}
using the fact that  variation of  determinant of an arbitrary matrix $X$ is equal to 
\begin{equation}
	\delta \sqrt{-\det X}=\frac{1}{2}\sqrt{-\det X}X^{ab}\delta X_{ba} \ ,
\end{equation}
and where we also  presume that the matter part of action depends on $g_{ab}$ only so that $\frac{\delta S_M}{\delta \Gamma^c_{ab}}=0$. 
Now from the requirement that the equations of motion follow from the stationary of the action we obtain following equations of motion for metric
\begin{eqnarray}\label{algrel}
&&	\sqrt{-\det\bA}\bAi^{ab}-\lambda \sqrt{-\det g}g^{ab}-\frac{\sqrt{-g}}{M_p^2M^2_{BI}}T^{ab}=0 \ ,  \nonumber \\
&& T^{ab}=-\frac{2}{\sqrt{-g}}\frac{\delta S_M}{\delta g_{ab}} \ .  \nonumber \\
\end{eqnarray}
 In case of the variation with respect to connection we should be more careful. First of all note that we have
\begin{equation}
R_{(ab)}=\partial_m \Gamma^m_{ab}+\Gamma^m_{mn}\Gamma^n_{ab}
-\Gamma^m_{an}\Gamma^n_{mb}-\frac{1}{2}(\partial_a \Gamma^m_{mb}+
\partial_b \Gamma^m_{ma}) \ . 
\end{equation}
Performing variation of this expression with respect to $\Gamma^a_{ab}$ and after some manipulation we obtain following equations of motion
\begin{eqnarray}\label{parbD}
	-\partial_m \bD^{ab}+\bD^{rs}\Gamma_{rs}^b\delta^a_m
	+\bD^{ab}\Gamma^p_{pm}-\bD^{an}\Gamma^b_{nm}-\bD^{nb}
	\Gamma^a_{nm}+
	\partial_n[\bD^{nb}\delta_m^a]=0 \ , \nonumber \\
\end{eqnarray}
where we introduced symmetric tensor density
\begin{equation}
	\bD^{ab}\equiv \sqrt{-\det\bA}\bAi^{ab} \ . 
\end{equation}
Note that the equation (\ref{parbD})  can be rewritten with the help of covariant derivatives  into the form
\begin{eqnarray}\label{eqmPal}
	-\nabla_m \bD^{ab}+\nabla_n\bD^{nb}\delta_m^a=0 \ , 
	\nonumber \\
\end{eqnarray}
where the covariant derivative of tensor density $\bD^{ab}$  of weight $-1$ is equal to 
\begin{equation}
	\nabla_m \bD^{ab}=\partial_m \bD^{ab}+\bD^{ap}\Gamma_{pm}^b+
	\Gamma^a_{mp}\bD^{pb}-\bD^{ab}\Gamma^p_{pm} \ . 
\end{equation}
Further, performing contraction between $a$ and $m$ (\ref{eqmPal}) we get
\begin{equation}
	\nabla_m \bD^{mb}=0 \ . 
\end{equation}
Inserting this result into (\ref{eqmPal}) 
we finally obtain
\begin{equation}
	\nabla_m \bD^{ab}=0 \ .
\end{equation}
Since $\bA$ is non-singular matrix the equation above can be rewritten into the form 
\begin{equation}
\nabla_m \bA^{ab}=0 \ . 
\end{equation}
 This equation implies that connection is Levi-Civita connection of the auxiliary metric where $\bA_{ab}$ can be fully determined by $g_{ab}$ and $T_{ab}$ as follows from (\ref{algrel}). Further, since connection has been solved by algebraic equations no additional physical degrees of freedom can be related to it. This is very attractive property of Born-Infeld gravity.
 
 In the next section we find covariant canonical form of BI gravity and we show that its structure allows us to find very easily corresponding Hamiltonian.
\section{Covariant Canonical Formalism of BI gravity}
\label{third}
Let us again write action for BI gravity
\begin{equation}\label{SBIcov}
	S=M_p^2M_{BI}^2
	\int d^4x\left(\sqrt{-\det \left(g_{ab}+\frac{1}{M_{BI}^2}R_{(ab)}\right)}
	-\lambda \sqrt{-\det g}\right)+S_M \ 
\end{equation}
from which we define corresponding covariant conjugate momenta. Note that they are defined 
 as variation  of the action with respect to all partial derivatives of given field. For example, in case of the action for scalar field $\mL_{\varphi}=-\frac{1}{2}g^{ab}\sqrt{-g}\partial_a\varphi\partial_b\varphi$ we get
\begin{equation}
	p^a_\varphi=\frac{\partial\mL_{\varphi}}{\partial(\partial_a\varphi)}=-\sqrt{-g}g^{ab}\partial_b\varphi \ . 
\end{equation}
This procedure can be easily generalized to more complicated tensor fields. In case of BI gravity action we have two dynamical fields $g_{ab}$ and $\Gamma^c_{ab}$ and hence corresponding conjugate covariant momenta are equal to
\begin{eqnarray}
	&&M^{abc}=\frac{\delta S}{\delta \partial_c g_{ab}}=0 \ , \nonumber \\
&&		\Pi^{abd}_{\quad c}=\frac{\partial S}{\partial(\partial_d \Gamma^c_{ab})}=
	\frac{1}{2}M_p^2\sqrt{-\det\bA}(\bA^{ab}\delta_c^d-\frac{1}{2}
	(\bA^{da}\delta_c^b+\bA^{db}\delta_c^a)) \ . 
	\nonumber \\
	\end{eqnarray}
To proceed further we use the fact that 
\begin{eqnarray}
&&	\Pi^{abd}_{ \quad c}(\Gamma^c_{am}\Gamma^m_{bd}+
	\Gamma^c_{bm}\Gamma^m_{ad})
=\frac{1}{2}
M_p^2\sqrt{-\det\bA}
\bA^{ab}[\Gamma^d_{am}\Gamma^m_{bd}-
\Gamma_{ab}^m\Gamma_{mn}^n]
\nonumber \\	
\end{eqnarray}
and also 
\begin{eqnarray}
&&	\Pi^{abd}_{\quad c}\Gamma^m_{ab}\Gamma_{md}^c=
	\frac{1}{2}M_p^2
	\sqrt{-\det\bA}\bA^{ab}[\Gamma^m_{ab}\Gamma_{mc}^c-
	\Gamma^m_{an}\Gamma^n_{mb}] \ ,
	\nonumber \\
&&	\Pi^{abd}_{ \quad c}M_{BI}^2
g_{ab}\delta^c_d=\frac{3}{2}M_p^2M_{BI}^2
	\sqrt{-\det\bA}\bA^{ab}g_{ab} \ . \nonumber \\
\end{eqnarray}
Then we can write 
\begin{eqnarray}
	\Pi^{abd}_{\quad c}\partial_d \Gamma^c_{ab}=
2M_p^2M_{BI}^2\sqrt{-\det\bA}+[\frac{1}{2}(	\Pi^{abd}_{ \quad c}(\Gamma^c_{am}\Gamma^m_{bd}+
\Gamma^c_{bm}\Gamma^m_{ad}
-\Gamma^m_{ab}\Gamma^c_{md})-\frac{M_{BI}^2}{3}\Pi^{abd}_{\quad c}g_{ab}\delta^c_d)]
\nonumber \\
		\end{eqnarray}
and hence covariant  Hamiltonian density is equal to 
\begin{eqnarray}
&&	\mH=\Pi^{abd}_{ \quad  c}\partial_d \Gamma_{ab}^c-\mL=
M_p^2M_{BI}^2\sqrt{-\det\bA}+\nonumber \\
&&+\frac{1}{2}[	\Pi^{abd}_{ \quad c}(\Gamma^c_{am}\Gamma^m_{bd}+
\Gamma^c_{bm}\Gamma^m_{ad}-\Gamma^m_{ab}\Gamma^c_{md})-\frac{2M_{BI}^2}{3}\Pi^{abd}_{\quad c}g_{ab}\delta^c_d]+
M_p^2M_{BI}^2\lambda \sqrt{-\det g} \ 
\nonumber \\
\end{eqnarray}
which cannot be final form of Hamiltonian density since it depends on $\bA$ that contains partial derivative of connection. In fact, we can express $\det\bA$ as function of canonical variables with the help of following manipulation
\begin{eqnarray}
	\Pi^{abc}_{ \quad c}=
	\frac{1}{2}M_p^2
	\sqrt{-\det \bA}(\bA^{ab}\delta^c_c-\bA^{ab})=
	\frac{3}{2}M_p^2
	\sqrt{-\det \bA}\bA^{ab} 
\end{eqnarray}
that allows us to express $\det\bA$ as function of canonical
variables 	
\begin{equation}	
\det\bA=\left(\frac{2}{3M_p^2}\right)^4	\det \Pi^{abc}_{\quad c} \ . 
\end{equation}
Using this result we obtain final form of covariant canonical Hamiltonian
\begin{eqnarray}
&&	\mH=\frac{4M^2_{BI}}{9M_p^2}\sqrt{-\det \Pi^{abc}_{\quad c}}+
\nonumber \\
&&	+[\frac{1}{2}	\Pi^{abd}_{ \quad c}(\Gamma^c_{am}\Gamma^m_{bd}+
	\Gamma^c_{bm}\Gamma^m_{ad}
-\Gamma^m_{ab}\Gamma^c_{md}	)-\frac{M_{BI}^2}{3}\Pi^{abd}_{\quad c}g_{ab}\delta^c_d]+
	M_p^2M_{BI}^2\lambda\sqrt{-\det g} \ . 
	\nonumber \\
\end{eqnarray}
Finally we consider matter contribution. For simplicity we represent matter
as collection of 
 $N$ scalar fields $\phi^A, A=1,\dots,N$ with the action 
\begin{equation}\label{Smat}
		S_M=-\frac{1}{2}\int d^4x\sqrt{-g}(g^{ab}\partial_a\phi^A\partial_b\phi^B K_{AB}+V(\phi)) \ ,
\end{equation}
where $K_{AB}=K_{BA}$ can generally depend on $\phi_A$. From matter action (\ref{Smat}) we obtain 
 momenta conjugate to $\partial_\alpha \phi^A$ to be equal to
\begin{equation}
p^{a}_{ A}=
-\sqrt{-g}g^{ab}\partial_b\phi^B K^{BA}
\end{equation}
so that contribution of matter to the Hamiltonian is equal to
\begin{eqnarray}
&&	\mH_{M}=p^a_A\partial_a \phi^A-\mL_{M}=
	-\sqrt{-g}g^{ab}\partial_a\phi^A
\partial_b\phi^B K_{AB}+
\frac{1}{2}\sqrt{-g}V(\phi)=\nonumber \\
&&=-\frac{1}{2\sqrt{-g}}p_A^\alpha g_{\alpha\beta}p^\beta_BK^{AB}+\frac{1}{2}\sqrt{-g}V \ . 
\nonumber \\
\end{eqnarray}
It is an attractive property of BI gravity in the covariant canonical formalism that the Hamiltonian density is function of conjugate momenta. This is different form the covariant formulation of Palatini form of Einstein-Hilbert action when the momentum conjugate to connection is related to canonical variables through the primary constraint in the theory. However we will show that similar form of the relation arises in BI gravity as a solution of equations of motion which will be studied below. 

The standard procedure how to determine equations of motion in the covariant canonical formalism is to consider canonical form of the action 
\begin{equation}
	S_{can}=\int d^4x (\Pi^{abc}_{\quad d}\partial_c\Gamma^d_{ab}+M^{abc}\partial_c g_{ab}+
	p^a_A\partial_a\phi^A-\mH-\mH_{M}+\lambda_{abc}M^{abc}) \ , 
\end{equation}
where we included primary constraints $M^{abc}\approx 0$ and where $\lambda_{abc}$ are corresponding Lagrange multipliers. In order to obtain equations of motion we perform variation of this action with respect to all canonical variables
\begin{eqnarray}
&&	\delta S=\int d^4x (\delta\Pi^{abc}_{\quad d}\partial_c\Gamma^d_{ab}
-\partial_c \Pi^{abc}_{ \quad d}\delta \Gamma^d_{ab}+	
\nonumber \\	
&&	+\delta M^{abc}\partial_c g_{ab}-\partial_c M^{abc}\delta g_{ab}
	+\delta p^a_A\partial_a\phi^A
	-\partial_a p^a_A\delta \phi^A
-\nonumber \\
&&-\frac{\delta \mH}{\delta M^{abc}}\delta M^{abc}-
\frac{\delta \mH}{\delta g_{ab}}\delta g_{ab}
-\frac{\delta \mH}{\delta \Pi^{abc}_{ \quad d}}\delta \Pi^{abc}_{\quad d}
-\frac{\delta \mH}{\delta \Gamma^d_{ab}}\delta \Gamma^d_{ab}
-\nonumber \\
&&-\frac{\delta\mH_{M}}{\delta g_{ab}}\delta g_{ab}
-\frac{\delta \mH_{M}}{\delta p^a_A}\delta p^a_A
-\frac{\delta \mH_{M}}{\delta \phi^A}\delta\phi^A
-\delta\lambda_{abc}M^{abc}-\lambda_{abc}\delta M^{abc}) \ , \nonumber \\
\end{eqnarray}
where we ignored boundary terms. Variations with respect to $g_{ab}$ and $M^{abc}$ together with $\lambda_{abc}$ imply following equations of motion
\begin{eqnarray}\label{eqg}
&&	\partial_c g_{ab}-\lambda_{abc}=0 \ , \quad M^{abc}=0 \ ,  \nonumber \\
&&	-\partial_c M^{abc}-\frac{\delta \mH_{matt}}{\delta g_{ab}}
	+\frac{M_{BI}^2}{3}\Pi^{abc}_{\quad c}-\frac{1}{2}M_p^2M_{BI}^2\lambda
	\sqrt{-\det g}g^{ab}=0 \ . \nonumber \\
	\end{eqnarray}
The first equation in (\ref{eqg}) can be solved for $\lambda_{abc}$ while using the condition  $M^{abc}=0$
the last equation in (\ref{eqg}) 	
	 implies an algebraic relation
\begin{equation}\label{alggen}
	\frac{\delta \mH_{M}}{\delta g_{ab}}
+\frac{M_{BI}^2}{3}\Pi^{abc}_{\quad c}-\frac{1}{2}\lambda M_p^2M_{BI}^2
	\sqrt{-\det g}g^{ab}=0 \ 
	\end{equation}	
	that could be interpreted as secondary constraints in the 
	standard terminology of systems with constraints.
	
Finally the variation of the action with respect to  $\Gamma^d_{ab}$ and $\Pi^{abd}_{\quad c}$ leads to the following equations of motion 
\begin{eqnarray}\label{eqGamma}
&&\partial_d\Gamma^c_{ab}-\frac{1}{2}(\Gamma^c_{am}\Gamma^m_{bd}+\Gamma^c_{bm}
\Gamma^m_{ad}-\frac{2M_{BI}^2}{3}g_{ab}\delta^c_d-\Gamma^m_{ab}\Gamma^c_{md})-\nonumber \\
&&-
\frac{2M_{BI}^2}{9M_p^2}\sqrt{-\det \Pi^{abc}_{\quad c}}\hat{\Pi}_{ab}\delta_d^c=0 \ , 
 \nonumber \\
&&\partial_c \Pi^{abc}_{ \ d}+\frac{1}{2}(\Pi^{amn}_{ \quad d}\Gamma^b_{mn}+
\Pi^{bmn}_{ \quad d}\Gamma^a_{mn}+\Pi^{mab}_{\quad n}\Gamma_{md}^n+
\nonumber \\
&&+\Pi^{mba}_{\quad n}\Gamma^n_{md}-\Pi^{abm}_{\quad n}\Gamma^n_{md}-
\frac{1}{2}\Pi^{mna}_{\quad d}\Gamma^b_{mn}-\frac{1}{2}
\Pi^{mnb}_{\quad d}\Gamma^a_{mn})=0 \ , 
\nonumber \\
\end{eqnarray}
where we introduced matrix $\hat{\Pi}_{ab}$ as inverse matrix to $\Pi^{abc}_{\quad c}$
\begin{equation}
	\Pi^{abc}_{\quad c}\hat{\Pi}_{bd}=\delta^a_d \ . 
\end{equation}

Now we proceed to the analysis of corresponding equations of motion when we start with the case of absence of matter.

\subsection{Absence of Matter}
Let us start with the situation when there is no matter so (\ref{alggen}) reduces into an equation
\begin{equation}\label{alggenno}
	\frac{1}{3}\Pi^{abc}_{\quad c}=\frac{\lambda}{2}M_p^2
	\sqrt{-\det g}g^{ab} \ . 
\end{equation}
Let us solve this equation with an ansatz
\begin{equation}\label{ansPi}
	\Pi^{abc}_{\quad d}=\sqrt{-g}(Kg^{ab}\delta^c_d+L(g^{ac}\delta_d^b+g^{bc}\delta_d^a) \ , 
\end{equation}
where $K$ and $L$ are unknown constants which will be determined by requirement of the consistency of equations of motion. First of all  from (\ref{alggenno}) we get  relation between $K$ and $L$ in the form
\begin{eqnarray}\label{LplusK}
2K+L=\frac{3}{4}M_p^2 \ .  
\end{eqnarray}
Inserting (\ref{ansPi}) into  the equation (\ref{eqGamma}) we obtain following result
\begin{eqnarray}\label{parGamma}
&&\sqrt{-g}K(\partial_d g^{ab}+g^{am}\Gamma_{md}^b+
\Gamma^a_{md}g^{mb})+\nonumber \\
&&+\sqrt{-\det g}g^{ab}(\frac{1}{2}K\partial_d g_{mn}g^{mn}+
(L-\frac{K}{2})\Gamma^m_{md})+\nonumber \\
&&+\sqrt{-g}(L\partial_c g^{ac}-\frac{1}{2}\partial_c g^{mn}g_{mn}g^{ac}+\frac{1}{2}(L-\frac{1}{2}K)g^{mn}
\Gamma_{mn}^a)\delta_d^b+\nonumber \\
&&+\sqrt{-g}(L\partial_c g^{bc}-L\frac{1}{2}\partial_c g^{mn}g_{mn}g^{bc}+\frac{1}{2}(L-\frac{1}{2}K)g^{mn}
\Gamma_{mn}^a)\delta_d^a=0\nonumber \\
\end{eqnarray}
Interestingly the equation (\ref{parGamma}) can be  solved by compatibility condition between 
metric and connection
\begin{equation}\label{compatcon}
	\partial_d g^{ab}+g^{am}\Gamma^b_{md}+\Gamma^a_{md}g^{mb}=0 \ 
\end{equation}
when the first bracket on (\ref{parGamma}) is trivially solved. Further, multiplying the equation (\ref{compatcon}) with $g_{ab}$ we obtain
\begin{equation}
	\partial_d g^{mn}g_{mn}=-2\Gamma^m_{md} 
\end{equation}
or equivalently 
\begin{equation}
	\partial_d g_{mn}g^{mn}=2\Gamma^m_{md} \ , 
\end{equation}
where we used  the fact that $\partial_d g_{mn}g^{nm}+g_{nm}\partial_dg^{nm}=0$. Inserting this result into the expression on the second line in (\ref{parGamma}) we obtain  
\begin{eqnarray}
	\frac{1}{2}K\partial_d g_{mn}g^{mn}+(L-\frac{K}{2})\Gamma^m_{md}=0
\nonumber \\
\end{eqnarray}
that can be solved for $L$ as 
\begin{equation}\label{LK}
 L=-\frac{K}{2} \ . 
\end{equation}
Finally we consider an expression in last two brackets  in (\ref{parGamma})
\begin{equation}
	L\partial_c g^{ac}+L\frac{1}{2}\partial_c g_{mn}g^{mn}g^{ac}+
	\frac{1}{2}(L-\frac{1}{2}K)g^{mn}\Gamma_{mn}^a \ . 
\end{equation}
Inserting (\ref{LK}) and also 
$\partial_d g_{mn}g^{mn}=2\Gamma_{md}^m$ into equation above we obtain
\begin{equation}
	L(\partial_c g^{ca}+\Gamma_{mc}^mg^{ca}+g^{mn}\Gamma_{mn}^a)=0
\end{equation}
that is obeyed as a consequence of equation (\ref{compatcon}). In other words we have shown that the equations of motion for $\Gamma$ in the vacuum are equivalent to the compatibility condition of metric. Note also that using (\ref{LK}) and (\ref{LplusK}) we obtain 
\begin{equation}
	\Pi^{abc}_{\quad d}=\frac{M_p^2}{2}\sqrt{-g}
(g^{ab}\delta^c_d-\frac{1}{2}(g^{ac}\delta^b_d+g^{bc}\delta^a_d)) \ . 
\end{equation}

Finally we return to the equations of motion for $\Pi^{abc}_{ \quad d}$ given on  the second line in 
(\ref{eqGamma}). Inserting (\ref{alggenno}) into it and using the fact that 
$ \hat{\Pi}_{ab}=\frac{2}{3\lambda M^2_p}\frac{1}{\sqrt{-g}}g_{ab}$ we obtain that 
it can be written in the form 
	\begin{eqnarray}\label{eqPino}
\partial_d\Gamma^c_{ab}-\frac{1}{2}(\Gamma^c_{am}\Gamma^m_{bd}+\Gamma^c_{bm}
\Gamma^m_{ad}-\frac{2M^2_{BI}}{3}g_{ab}\delta^c_d-\Gamma^m_{ab}\Gamma^c_{md})-
\frac{1}{3}\lambda M^2_{BI}g_{ab}\delta_d^c=0 \ . 
\nonumber \\		
\end{eqnarray}
Now contraction between $d$ and $c$ indices gives 
\begin{eqnarray}\label{partdGamma}
	\partial_d \Gamma^d_{ab}-\frac{1}{2}
	(\Gamma^d_{am}\Gamma^m_{bd}+\Gamma^d_{bm}\Gamma_{ad}^m
-\frac{8M^2_{BI}}{3}g_{ab}-\Gamma^m_{ab}\Gamma^d_{dm})-\frac{4M^2_{BI}}{3}\lambda g_{ab}=0 \ .  \nonumber \\	
\end{eqnarray}
On the other hand contraction of  $c$ and $b$ indices in (\ref{eqPino}) leads to
\begin{equation}\label{partdGamma2}
	\partial_a \Gamma^m_{bm}-\frac{1}{2}(
\Gamma^n_{bm}\Gamma^m_{na}+\Gamma^n_{nm}\Gamma^m_{ab}-
\frac{2M^2_{BI}}{3}g_{ba}-\Gamma^m_{bn}\Gamma^n_{ma})-\frac{M^2_{BI}}{3}\lambda g_{ba}=0 \ . 
\end{equation}
If we combine (\ref{partdGamma}) together with (\ref{partdGamma2}) we obtain 
\begin{eqnarray}\label{partGammafin}
&&	\partial_m \Gamma^m_{ab}-\partial_a \Gamma^m_{bm}
+\Gamma^m_{ab}\Gamma^d_{dm}-
\Gamma^d_{am}\Gamma^m_{bd}+	
M_{BI}^2(1-\lambda)g_{ab}=
	\nonumber \\
&&=R_{ab}+M_{BI}^2(1-\lambda)g_{ab}=0 \nonumber \\	
\end{eqnarray}
which are exactly Einstein equations of motion in absence of matter and with a presence of cosmological constant $\Lambda=M_{BI}^2(1-\lambda)$. To see this note that gravitational equations of motion that follow from Einstein-Hilbert action $S=M_p^2\int d^4x\sqrt{-g}(R+2\Lambda)$ have the form 
\begin{equation}\label{eqRab}
	R_{ab}-\frac{1}{2}g_{ab}R-\Lambda g_{ab}=0 \ . 
\end{equation}
Contracting this equation with $g^{ab}$ we get
\begin{equation}
	-R=4\Lambda 
\end{equation}
which inserting back to (\ref{eqRab}) gives 
\begin{equation}
	R_{ab}+\Lambda g_{ab}=0
\end{equation}
which agrees with (\ref{partGammafin}). 

We have shown that covariant canonical formulation of BI gravity nicely reproduces equations of 
of motion of Palatini gravity in the absence of matter. In the next section we generalize this analysis to the case of non-zero matter contribution. 


\subsection{Including matter}
The situation is more complicated when the matter is included. Let us start with the second equation 
in (\ref{eqGamma}). Performing the same manipulation as we did in the end of previous section we obtain 
\begin{equation}\label{Rabmatt}
R_{ab}+M_{BI}^2g_{ab}-\frac{2M_{BI}^2}{3M_p^2}
\sqrt{-\det \Pi^{abc}_{\quad c}}
\hat{\Pi}_{ab}=0 \ .  
\end{equation}
On the other hand note that $\Pi^{abc}_{\quad c}$ obeys the equation 
(\ref{alggen}) that we rewrite into the form 
\begin{eqnarray}\label{Pimatt}
&&	\Pi^{abc}_{\quad c}=\frac{3}{M_{BI}^2}(-\frac{\delta \mH}{\delta g^{ab}}
+\frac{1}{2}\lambda M_p^2M_{BI}^2\sqrt{-g}g^{ab})=\nonumber \\
&&=\frac{3M_p^2}{2}\sqrt{-g}(
\frac{1}{M_p^2M_{BI}^2}T^{ab}+\lambda g^{ab})\equiv \frac{3M^2_p}{2}
\sqrt{-g}\Omega^{ab} \ ,
	\end{eqnarray}
	where
	\begin{equation}
\Omega^{ab}=\frac{1}{M_p^2M_{BI}^2}T^{ab}+\lambda g^{ab} \ , \quad 		T^{ab}=-\frac{2}{\sqrt{-g}}\frac{\delta \mH_{M}}{\delta g_{ab}} \ . 
	\end{equation}
Using (\ref{Pimatt}) we can rewrite (\ref{Rabmatt}) into the form 
\begin{equation}\label{Rabgen}
	R_{ab}+M_{BI}^2g_{ab}-\sqrt{-g}M_{BI}^2\sqrt{-\det \Omega}\hat{\Omega}_{ab}=0 \ , 
\end{equation}
where $\hat{\Omega}_{ab}$ is inverse matrix to $\Omega^{ab}$
\begin{equation}
	\Omega_{ab}\hat{\Omega}^{bc}=\delta_a^c \ . 
\end{equation}
It is instructive to analyze this equation in the limit when $T^{ab}\ll M_{BI}^2$\footnote{In fact, it is easy to see that BI gravity reduces to the Einstein-Hilbert form of the action in the limit $M_{BI}^2\rightarrow \infty$ or equivalently in the limit when the curvature is small with respect to $M_{BI}^2$.}. In case when $\frac{T^{ab}}{M_p^2 M_{BI}^2}\ll 1$  we can write
\begin{equation}
	\Omega^{ab}=\lambda g^{ac}(\delta_c^b+\frac{1}{\lambda M_p^2M_{BI}^2}g_{cd}T^{db}) \ , 
	\quad \hat{\Omega}_{ab}=\frac{1}{\lambda}g_{ab}-\frac{1}{\lambda^2 M_p^2M_{BI}^2}T_{ab} \ , 
	\quad T_{ab}=g_{ac}g_{bd}T^{cd} \ . 
\end{equation}
In this approximation (\ref{Rabgen}) reduces into 
\begin{eqnarray}\label{Rmat1}
	R_{ab}+M_{BI}^2(1-\lambda)-\frac{1}{2 M_p^2}g_{cd}T^{dc}+\frac{1}{ M_p^2}T_{ab}=0
\nonumber \\	
\end{eqnarray}
which are equivalent to standard Einstein equations in the presence of matter and with the cosmological constant $\Lambda=(1-\lambda)M_{BI}^2$ 
\begin{equation}\label{Rmat}
	R_{ab}-\frac{1}{2}Rg_{ab}-g_{ab}\Lambda+\frac{1}{M_p^2}T_{ab}=0 \ . 
\end{equation}
Again, from this equation we can express $R$ as 
\begin{equation}
	R=\frac{T}{M_p^2}-4\Lambda \ , T\equiv T_{ab}g^{ab} 	
\end{equation}
and inserting back to (\ref{Rmat}) we obtain (\ref{Rmat1}) which demonstrates an equivalence of the BI gravity with ordinary gravity in the limit of large $M^2_{BI}$. Note also that there is significant deviation from Einstein theory in case when $M_{BI}^2$ is finite which makes BI inspired gravity very interesting generalization of standard general relativity. 

Let us proceed with the analysis of covariant canonical form of gravity with remaining equations of motion. 
	Let us introduce an auxiliary metric $\hat{g}^{ab}$ through the relation 
\begin{equation}\label{Pihatg}
	\Pi^{abc}_{\quad c}=\sqrt{-\det\hat{g}}\hat{g}^{ab}
\end{equation}
so that
\begin{equation}
	\sqrt{-\det\hat{g}}\hat{g}^{ab}=
\frac{3M_p^2}{2}\sqrt{-g}\Omega^{ab} \ 
\end{equation}
that allows us to find $\hat{g}$ as function of $g$ and $T$. Explicitly, from previous equation we obtain  
\begin{equation}
	\det \hat{g}_{ab}=\left(\frac{3}{2}M_p^2\right)^4\det g^2\det \Omega^{ab} \  
\end{equation}
so that
\begin{equation}
	\hat{g}^{ab}=\frac{3M_p^2\sqrt{-g}}{2\sqrt{-\hat{g}}}
	\Omega^{ab} \ . 
\end{equation}
Since (\ref{Pihatg}) has formally the same form as the relation between $\Pi^{abc}_{\quad c}$ in the absence of matter we can again propose an ansatz 
\begin{equation}
\Pi^{abc}_{\quad d}=K\sqrt{-\hat{g}}(\hat{g}^{ab}\delta^c_d+
L(\hat{g}^{ac}\delta^b_d+
\hat{g}^{bc}\delta^a_d)) \ . 
\end{equation}
Then the first equation in (\ref{eqGamma}) again implies compatibility condition  between an auxiliary metric $\hat{g}^{ab}$ and $\Gamma^a_{bc}$ and also the relation $L=-\frac{K}{2}$. This result together with (\ref{Rabgen}) shows that covariant canonical formalism of BI gravity nicely reproduces Lagrangian equations of motion. On the other hand it is important to stress that when we include matter this theory significantly deviates from standard general relativity.

{\bf Acknowledgment:}

This work  is supported by the grant “Dualitites and higher order derivatives” (GA23-06498S) from the Czech Science Foundation (GACR).


\end{document}